# Stochastic Unit Commitment in Electricity-Gas Coupled Integrated Energy Systems based on Modified Progressive Hedging

Haizhou Liu, *Student Member*, *IEEE*, Xinwei Shen, *Member*, *IEEE*,
Qinglai Guo, *Senior Member*, *IEEE*, Hongbin Sun, *Fellow*, *IEEE*,
Wenzhi Zhao and Xinyi Zhao

***Abstract*—The increasing number of gas-fired units has significantly intensified the coupling between power and gas networks. Traditionally, the nonlinearity and nonconvexity in gas flow equations, together with renewable-induced stochasticity, result in a computationally expensive model for unit commitment in electricity-gas coupled integrated energy systems (IES). To accelerate stochastic day-ahead scheduling, we applied and modified Progressive Hedging (PH), a heuristic approach that can be computed in parallel to yield scenario-independent unit commitment. By applying a termination and enumeration technique, the modified PH algorithm saves considerable computational time, especially when the unit production prices are similar for all generators, and when the scale of IES is large. Moreover, an adapted second-order cone relaxation (SOCR) is utilized to tackle the nonconvex gas flow equation. Case studies are performed on the IEEE 24-bus system/Belgium 20-node gas system and the IEEE 118-bus system/Belgium 20-node gas system. The computational efficiency when employing PH is 188 times that of commercial software, even outperforming Benders Decomposition. Meanwhile, the gap between the PH algorithm and the benchmark is less than 0.01% in both IES systems, which proves that the solution produced by PH reaches acceptable optimality in this stochastic UC problem.**

## NOMENCLATURE

### A. Indexes and Sets

| | |
|---|---|
| $i \in \mathcal{G}$ | Index of gas-fired units. |
| $i \in \mathcal{C}$ | Index of coal-fired units. |
| $g \in \mathcal{GW}$ | Index of gas wells. |
| $sc \in \mathcal{SC}$ | Index of typical scenarios. |
| $l \in \mathcal{L}$ | Index of electric buses. |
| $n \in \mathcal{N}$ | Index of gas nodes. |
| $s \in \mathcal{S}$ | Index of gas storage. |
| $w \in \mathcal{W}$ | Index of wind farms. |
| $t \in \mathcal{T}$ | Hourly time periods for the following day. |
| $(a,b) \in \mathcal{BR}$ | Transmission lines from bus $a$ to $b$. |
| $(c,d) \in \mathcal{PL}$ | Gas pipelines from gas node $c$ to $d$. |
| $(q,j) \in \mathcal{PC}$ | Compressors; $q$ and $j$ denote the starting and terminal nodes of a compressor, respectively. |

### B. Parameters

| | |
|---|---|
| $C_i^{PD}$ | Unit production cost of generators (M\$/GW). |
| $C_g^{PG}$ | Unit cost of gas generation (M\$/GW). |
| $P^{sc}$ | Probability of occurrence for a typical wind output scenario $sc$. |
| $C_l^{NP}$ | Unit cost for non-served power at bus $l$ (M\$/GWh). |
| $C_n^{NG}$ | Unit cost for non-served gas at node $n$ (M\$/MSm$^3$). |
| $C_s^{S}$ | Unit cost for storage $s$ (M\$/MSm$^3$). |
| $C_w^{WC}$ | Unit cost for wind curtailment at wind farm $w$ (M\$/GW). |
| $RU_i$, $RD_i$ | Ramp up/down limits for generator $i$ (GW). |
| $\underline{P}_i, \overline{P}_i$ | Minimum and maximum power output of unit $i$ (GW). |
| $PF_{ab}$ | Maximum power flow of branch $ab$ (GW). |
| $L_{l,t}^{P}$ | Power demand at bus $l$ and time $t$ (GW). |
| $X_{ab}$ | Inductance of branch $(a,b)$ (p. u.). |
| $\underline{W}_g, \overline{W}_g$ | Minimum and maximum production for gas well $g$ (MSm$^3$). |
| $\underline{S}_s, \overline{S}_s$ | Minimum and maximum storage levels for storage $s$ (MSm$^3$). |
| $\overline{WR}_s, \overline{IS}_s$ | Withdrawal and injection limits for storage $s$ (MSm$^3$/h). |
| $D_{cd}$ | Diameter of pipeline $(c,d)$ (m). |
| $L_{cd}$ | Length of pipeline $(c,d)$ (m). |

This work is supported by the National Key Research and Development Program SQ2019YFE020392 and the Science, Technology and Innovation Commission of Shenzhen Municipality (No.JCYJ20170411152331932).

The authors, except for Wenzhi Zhao, are with Tsinghua-Berkeley Shenzhen Institute, Shenzhen International Graduate School, Tsinghua University, Shenzhen, 518055, China. In addition, Prof. Qinglai Guo, Prof. Hongbin Sun and Wenzhi Zhao are with Department of Electrical Engineering, Tsinghua University, 100084, Beijing, China. (Corresponding author: Dr. Xinwei Shen, sxw.tbsi@sz.tsinghua.edu.cn; Prof. Hongbin Sun, shb@tsinghua.edu.cn).

| | |
|---|---|
| $\rho$ | Density of natural gas (kg/m$^3$). |
| $F$ | Friction coefficient. |
| $R$ | Specific gas constant (m$^3$·bar·K$^{-1}$·kg$^{-1}$). |
| $T$ | Temperature of the gas system (K). |
| $Z$ | Average compressibility factor of gas. |
| $GTP_i$ | Gas-to-power conversion efficiency for gas turbine $i$ (MSm$^3$/GWh). |
| $\underline{\Pi}_n, \overline{\Pi}_n$ | Minimum and maximum pressure for gas node $n$ (bar). |
| $CM_{qj}$ | Compression factor, for compressor located between gas nodes $q$ and $j$. |

**C. Variables**

| | |
|---|---|
| $pd^{sc}_{i,t}$ | Power production of unit $i$ at time $t$ and scenario $sc$ (GW). |
| $pg^{sc}_{g,t}$ | Gas production of gas well $g$ at time $t$ and scenario $sc$ (MSm$^3$/h). |
| $np^{sc}_{l,t}$ | Non-served power at bus $l$ at time $t$ and scenario $sc$ (GWh). |
| $ng^{sc}_{n,t}$ | Non-served gas at node $n$ at time $t$ (MSm$^3$). |
| $sto^{sc}_{s,t}, sti^{sc}_{s,t}$ | Gas withdrawn from/injected into storage $s$ at time $t$ and scenario $sc$ (MSm$^3$/h). |
| $wc^{sc}_{w,t}$ | Wind curtailment of wind farm $w$ at time $t$ and scenario $sc$ (GW). |
| $c_{i,t}$ | Unit commitment for unit $i$ at time $t$. |
| $ru^{sc}_{i,t}, rd^{sc}_{i,t}$ | Up/down ramping for unit $i$ at time $t$ and scenario $sc$ (GW). |
| $pf^{sc}_{ab,t}$ | Power flow for branch ($a$,$b$) at time $t$ (GW). |
| $\theta^{sc}_{a,t}, \theta^{sc}_{b,t}$ | Phase angle at bus $a$, $b$ at time $t$ and scenario $sc$. |
| $sl^{sc}_{s,t}$ | Storage level of storage $s$ at time $t$ and scenario $sc$ (MSm$^3$). |
| $\pi^{sc}_{c,t}, \pi^{sc}_{d,t}$ | Pressure at gas node $c$, $d$ at time $t$ and scenario $sc$ (bar). |
| $\pi^{sc}_{q,t}, \pi^{sc}_{j,t}$ | Pressure at gas node $q$, $j$ at time $t$ and scenario $sc$ (bar). |
| $\overline{\pi}^{sc}_{cd,t}$ | Average pressure of gas node $c$ and $d$ at time $t$ and scenario $sc$ (bar). |
| $m^{sc}_{cd,t}$ | Amount of gas stored in pipeline ($c$,$d$) at time $t$ and scenario $sc$ (MSm$^3$). |
| $gfo^{sc}_{cd,t}, gfi^{sc}_{cd,t}$ | Gas flow out of and into pipeline ($c$,$d$) at time $t$ and scenario $sc$ (MSm$^3$/h). |
| $gf^{sc}_{cd,t}$ | Average gas flow of pipeline ($c$,$d$) at time $t$ and scenario $sc$ (MSm$^3$/h). |

# I. INTRODUCTION

DUE to the increasing demand for clean and efficient energy resources, electricity-gas coupled networks have garnered considerable interest in recent decades, as an example of integrated energy systems (IESs). Gas-fired units are pollutant-free in terms of sulfur or nitrogen oxides, and highly efficient in energy conversion compared to coal-fired units [1]. It has been predicted that power generated by gas-fired units will increase by 230% by 2030 [2][3], and natural gas consumption would take up 28% of all energy demand by then [4]. Moreover, the refinement of Power-to-Gas (PtG) techniques [5] has strengthened the interaction of power and gas systems. In addition to environmental benefits such as carbon consumption, PtG units enable large-scale energy storage, which would further improve the flexibility of IESs.

Regarding coupled IES systems, reliability and security are two of the most crucial topics. Generally, sources of uncertainty in IESs include renewable generation fluctuation [6], load variation [7], N-1 contingencies [8] and malicious attacks [9]. Gas systems are generally considered an ideal choice for accommodating renewable energy, such as wind and solar energy. The pipelines serve as flexible storage that draws in/out gases by self-adjusting nodal pressures, outperforming power reserves in terms of cost and stability. However, eco-friendly renewables introduce operational uncertainties that would jeopardize the coupled systems. [10] introduced the concept of distributed slack nodes to analyze gas flow when wind power is included. [11] established an EGTran model that aims to solve for the wind-penetrated stochastic unit commitment (UC). However, the UCs vary among different scenarios, and the gas dynamics are not considered, leading to a less instructive conclusion for operation in the following day.

When addressing renewable-induced uncertainties, two branches emerge: robust optimization (RO) and stochastic optimization (SO). RO recasts a deterministic model over an uncertainty set with upper and lower bounds, followed by identifying and optimizing the worst-case scenarios [12]. Although fewer data are required, RO is still limited due to its overconservativeness. On the other hand, SO resolves a problem by selecting typical scenarios and optimizing the resulting joint model, thus making it popular in research on UC [11][13][14]. One of the problems in SO is the explosion of scenarios, which significantly complicates the model. Stochastic models are normally accompanied by algorithms for acceleration. One of the most famous algorithms is the Benders Decomposition [15], the acceleration performance of which has been confirmed.

However, these algorithms are not specifically designed for stochastic scenarios. Scenario-independent variables (SIVs) are still correlated among scenarios, which make the computation unable to be performed in parallel. In light of this, Progressive Hedging (PH) is introduced as a scenario-based decomposition method. First proposed by [16], PH is known as a distributed technique that keeps solving each scenario individually. Convergence of this article is also strictly proven [17]. PH largely resembles the Alternating Direction Method of Multipliers (ADMM) as a heuristic and distributed algorithm [18], but relaxes non-anticipativity constraints instead of linking constraints, therefore not requiring auxiliary functions or matrix inversion [19]. Moreover, PH is inherently supportive of parallel computing, which may further facilitate the optimization due to the uniformity of the decomposed scenarios [17]. The PH algorithm has been applied to the stochastic UC of power systems [20][21], and the optimality gap has been confirmed to tighten toward zero with a larger number of iterations [22][23]. However, the exploration of this algorithm in the operation of IESs has been rare. The introduction of PH can be meaningful, especially in a natural gas system where each scenario contains hundreds of nonlinear and nonconvex gas flow equations.

To model an electricity-gas coupled IES, the steady-state formulation has been widely adopted due to its simplicity [24][25]. Nevertheless, in reality, pipelines are equipped with the flexibility to adjust their gas storage level by changing the nodal pressures. Hence, the dynamics of pipelines remain a widely investigated topic, especially for short-term operation where pipeline storage makes a difference in economic dispatch. Liu et al. first established a gas transmission model that describes a slow transient through partial differential equations and algebraic equations [26]. Correa-Posada et al. further simplified the dynamic model into a mixed-integer linear programming (MILP) problem. They formed a linear linepack equation and gas flow equation by considering the gas travel velocity and compressibility [27]. Although less accurate, the linear linepack model is still favorable in short-term operation, as it recovers most of the dynamics.

To address the only nonconvex gas flow constraints, various piecewise linearization techniques (summarized in [28]) have been proposed. These algorithms can mostly recover a bilinear equation, yet a large number of segments are required to guarantee the exactness. This would introduce hundreds of binary variables, and is therefore inefficient in computation. Second-order cone relaxation (SOCR), as an alternative in addressing gas flow, has been extensively investigated in the past 10 years. It does not require additional binary variables, which significantly speeds up the calculation. Cone reformulation techniques including Taylor expansion-based linearization [29], sequential-cones [30] and convex envelopes [31], can further reduce the error to negligible levels. The ex-post recovery issues generated from SOCR are addressed in [32], [33] and [34].

In view of the above, this article applies and modifies the PH algorithm to solve stochastic UC problems in IESs, with a dynamic model and an iteration-free SOCR modeling the gas flow. The contribution of this article is therefore two-fold:

1) Propose a stochastic UC model for electricity-gas coupled IESs, where SOCR-based dynamics of the natural gas system are considered. To the best of our knowledge, this topic has not been explored due to problems in computation efficiency. Additionally, we use an iteration-free SOCR technique to model the gas flow, in which a linear constraint is proposed to compensate for the inexactness in relaxation.

2) Accelerate solving the stochastic UC problem based on a modified PH algorithm, which has outperformed commercial software and even the Benders Decomposition. In cases where the network is large, or the unit production prices are disparate for either generator, traditional Progressive Hedging (TPH) is sufficient. However, when the scale of the IES increases, or when unit power production prices are similar, we develop modified Progressive Hedging (MPH) to further reduce the computational burden, in which the proposed termination criteria control the number of iterations required for convergence.

The remainder of this paper is structured as follows: Section II formulates the stochastic UC model, with the generation of wind output scenarios. Section III illustrates the TPH/MPH algorithm and SOCR method for improving the performance of optimization. Section IV tests the novel algorithms on medium and large-scale IES systems to evaluate the algorithm's performance, and compares them with a benchmark method and Benders Decomposition. Section V concludes the article.

## II. Model Formulation

### A. Objective Function

The stochastic UC model for IESs, adapted from [27], considers multiple scenarios of wind power. The objective is to minimize the total expected cost incurred within the coupled system, over the time horizon $t\in\mathcal{T}$ and scenarios with superscript $sc\in\mathcal{SC}$

$$\min \sum_{sc\in\mathcal{SC}} P^{sc} \sum_{t\in\mathcal{T}} \Big[ \Big( \sum_{i\in\mathcal{G}\cup\mathcal{C}} C_i^{PD} pd_{i,t}^{sc} \Big) + \Big( \sum_{g\in\mathcal{GW}} C_g^{PG} pg_{g,t}^{sc} + \sum_{s\in\mathcal{S}} C_s^{S} sto_{s,t}^{sc} \Big) + \Big( \sum_{l\in\mathcal{L}} C_l^{NP} np_{l,t}^{sc} + \sum_{n\in\mathcal{N}} C_n^{NG} ng_{n,t}^{sc} + \sum_{w\in\mathcal{W}} C_w^{WC} wc_{w,t}^{sc} \Big) \Big] \quad (1)$$

where $\sum_{i\in\mathcal{G}\cup\mathcal{C}} C_i^{PD} pd_{i,t}^{sc}$ calculates the generation cost of all gas- and coal-fired generators, and $\sum_{g\in\mathcal{GW}} C_g^{PG} pg_{g,t}^{sc} + \sum_{s\in\mathcal{S}} C_s^{S} sto_{s,t}^{sc}$ considers the additional costs within the natural gas system, respectively gas production from wells and withdrawal from storage. Finally, non-served power for power and gas, together with curtailment of wind power, are penalized.

### B. Power System Constraints

The output power from generators is constrained by unit commitment, unit capacity limits (2) and up/down ramping limits (3). Note that the unit commitment $c_{i,t}$'s are the only scenario-independent variables.

$$c_{i,t} \underline{P}_i \le pd_{i,t}^{sc} \le c_{i,t} \overline{P}_i, \forall i \in \mathcal{G}\cup\mathcal{C} \quad (2)$$

$$-RD_i \le pd_{i,t}^{sc} - pd_{i,t-1}^{sc} \le RU_i, \forall i \in \mathcal{G}\cup\mathcal{C} \quad (3)$$

For power transmission, a lossless DC model is employed for simplicity in (4). Constraint (5) restricts the power flow of each transmission line by its capacity. Constraint (6) outlines the nodal balance at each bus, which involves power generation, transmission, consumption and VOLL.

$$pf_{ab,t}^{sc} = \left(\theta_{a,t}^{sc} - \theta_{b,t}^{sc}\right) / X_{ab}, \forall (a,b) \in \mathcal{BR} \quad (4)$$

$$-PF_{ab} \le pf_{ab,t}^{sc} \le PF_{ab}, \forall (a,b) \in \mathcal{BR} \quad (5)$$

$$pd_{l,t}^{sc} + \sum_{b=l} pf_{ab,t}^{sc} - \sum_{a'=l} pf_{a'b',t}^{sc} + np_{l,t}^{sc} = L_{l,t}^{P}, \forall l \in \mathcal{BR} \quad (6)$$

### C. Natural Gas System Constraints

A typical natural gas system consists of gas wells, storage, compressors and pipelines. Constraint (7) describes the production limit of natural gas wells. Constraints for storage include storage capacities (8), injection/withdrawal upper limits (9), and the variation in storage levels with time (10). For compressors, Constraint (11) describes the nodal pressure requirements between the starting and ending gas nodes. For simplicity, we assume that compressors do not consume energy.

$$\underline{W}_g \le pg_{g,t}^{sc} \le \overline{W}_g, \forall g \in \mathcal{GW} \tag{7}$$

$$\underline{S}_s \le sl_{s,t}^{sc} \le \overline{S}_s, \forall s \in \mathcal{S} \tag{8}$$

$$sto_{s,t}^{out} \le \overline{WR}_s, sti_{s,t}^{sc} \le \overline{IR}_s, \forall s \in \mathcal{S} \tag{9}$$

$$sl_{s,t+1}^{sc} = sl_{s,t}^{sc} + sto_{s,t}^{sc} - sti_{s,t}^{sc}, \forall s \in \mathcal{S} \tag{10}$$

$$\pi_q^{sc} \le \pi_j^{sc} \le CM_{qj}\pi_q^{sc}, \forall (q,j) \in \mathcal{PC} \tag{11}$$

For pipelines, we apply a quasi-dynamic model following [27]. Constraint (12) denotes the equation of mass within a pipeline, and (13) denotes its variation with time. The average gas flow, defined as the mean of *gfo* and *gfi* (14), is linked with pressure difference by the Weymouth equation (15).

$$\overline{\pi}_{cd,t}^{sc} \cdot 0.78 D_{cd}^2 L_{cd} = \rho m_{cd,t}^{sc} RTZ, \ \forall (c,d) \in \mathcal{PL} \tag{12}$$

$$m_{cd,t+1}^{sc} = m_{cd,t}^{sc} + gfi_{cd,t}^{sc} - gfo_{cd,t}^{sc}, \forall (c,d) \in \mathcal{PL} \tag{13}$$

$$gf_{cd,t}^{sc} = \left(gfo_{cd,t}^{sc} + gfi_{cd,t}^{sc}\right)/2, \forall (c,d) \in \mathcal{PL} \tag{14}$$

$$gf_{cd}^{sc}\left|gf_{cd}^{sc}\right| = 0.617 D_{cd}^5 \frac{\pi_{c,t}^{sc\,2} - \pi_{d,t}^{sc\,2}}{L_{cd} FRTZ\rho^2}, \forall (c,d) \in \mathcal{PL} \tag{15}$$

As shown in (16), each node in the gas system is subject to upper/lower pressure boundaries. Constraint (17) states the nodal balance within the natural gas system. An additional constraint is imposed to refrain the minimum pipeline storage by the end of the day (18)

$$\underline{\Pi}_n \le \pi_{n,t}^{sc} \le \overline{\Pi}_n, \forall n \in \mathcal{N} \tag{16}$$

$$\begin{aligned} &\sum_{d'=n} gfo_{c'd',t}^{sc} - \sum_{c=n} gfi_{cd,t}^{sc} + \sum_{s=n}\left(sto_{s,t}^{sc} - sti_{s,t}^{sc}\right) - GTP_i \times \sum_{i=n} pd_{i,t}^{sc} \\ &+ ng_{n,t}^{sc} = L_{n,t}^G, \forall n \in \mathcal{N}, (c,d) \in \mathcal{PL}, (c',d') \in \mathcal{PL} \end{aligned} \tag{17}$$

$$\sum_{(c,d)} m_{cd,t=24}^{sc} = \sum_{(c,d)} m_{cd,t=0}^{sc}, \forall (c,d) \in \mathcal{PL} \tag{18}$$

### D. Model of Wind Power Generation

Generating characteristic output curves of wind turbines is the key to constructing accurate stochastic scenarios. Herein the real-time windspeed $v_r(t)$ is viewed as the summation of both the forecasted windspeed $v_f(t)$ and the prediction error $v_e(t)$

$$v_r(t) = v_f(t) + v_e(t) \tag{19}$$

Considering the difficulty in predicting windspeed from only historical records, $v_f(t)$ is obtained by importing a 24-hour windspeed curve, from the National Data Buoy Center (NDBC) dataset [35]. $v_e(t)$ is generated through the auto-regressive moving average (ARMA) series. Here a simplified ARMA(1,1) model, in which the current error is only correlated with error in the past hour with additional noise, is chosen in accordance with references [6] and [36] as follows:

$$v_e(t) = \alpha v_e(t-1) + \beta\xi(t-1) + \xi(t) \tag{20}$$

where $\xi(t)$ is a random variable following a Gaussian distribution, with a mean of 0 and a standard deviation of $\sigma$. Random scenarios can therefore be generated from (20).

Subsequently, k-means clustering is applied to extract typical scenarios of windspeed. k-means will inevitably lead to deviations in the distribution of windspeed, but the computational burden can be greatly relieved. Otherwise the optimality gap would hardly converge even for decomposition algorithms. The reduced scenarios are finally mapped to the power output through a typical power curve [37].

## III. Algorithms: TPH/MPH and SOCR

### A. Traditional/Modified Progressive Hedging

As stated in Section I, a modified Progressive Hedging approach is applied to address the stochastic nature of wind turbines. For the sake of simplicity, the stochastic model is expressed in a compact form as follows:

$$\begin{aligned} &\text{minimize} \quad \boldsymbol{\lambda}_0^T \boldsymbol{c} + \sum_{sc} P_{sc} \boldsymbol{\lambda}_{sc}^T \boldsymbol{x}_{sc} \\ &\text{subject to} \quad (\boldsymbol{c}, \boldsymbol{x}_{sc}) \in \mathcal{Q}_{sc} \end{aligned} \tag{21}$$

where $\boldsymbol{c}$ denotes UC variables, the vectorized SIVs. All other variables are categorized by scenario, and are subsequently vectorized into $\boldsymbol{x}_{sc}$. $\mathcal{Q}_{sc}$ denotes the feasible region for each scenario, which is the feasible set defined by (2)-(18). $\boldsymbol{\lambda}_0$ and $\boldsymbol{\lambda}_{sc}$ are linear coefficients of the corresponding variables.

**Algorithm 1: TPH/MPH**

**Parallel For** each scenario
  $\boldsymbol{c}_{sc}^0, \boldsymbol{x}_{sc}^0 = \underset{\boldsymbol{c},\boldsymbol{x}_{sc}}{\text{argmin}} \quad \boldsymbol{\lambda}_0^T \boldsymbol{c} + \boldsymbol{\lambda}_{sc}^T \boldsymbol{x}_{sc}$
    subject to $(\boldsymbol{c}, \boldsymbol{x}_{sc}) \in \mathcal{Q}_{sc}$
**End Parallel**
$\overline{\boldsymbol{c}}^0 \leftarrow \sum_{sc} P_{sc} \boldsymbol{c}_{sc}^0$
$iter \leftarrow 0, \boldsymbol{\rho}_{sc}^0 \leftarrow \kappa\left(\boldsymbol{c}_{sc}^0 - \overline{\boldsymbol{c}}^0\right)$
**Do**
  $iter \leftarrow iter + 1$
  **Parallel For** each scenario
    $\boldsymbol{c}_{sc}^{iter}, \boldsymbol{x}_{sc}^{iter} = \underset{\boldsymbol{c},\boldsymbol{x}_{sc}}{\text{argmin}} \quad \boldsymbol{\lambda}_0^T \boldsymbol{c} + \boldsymbol{\lambda}_{sc}^T \boldsymbol{x}_{sc} + \boldsymbol{\rho}_{sc}^{iter-1\,T} \boldsymbol{c} + \frac{\kappa}{2}\left\|\boldsymbol{c} - \overline{\boldsymbol{c}}^{iter-1}\right\|_2^2$
      subject to $(\boldsymbol{c}, \boldsymbol{x}_{sc}) \in \mathcal{Q}_{sc}$
    $\boldsymbol{\rho}_{sc}^{iter} \leftarrow \boldsymbol{\rho}_{sc}^{iter} + \kappa\left(\boldsymbol{c}_{sc} - \overline{\boldsymbol{c}}\right)$
  **End Parallel**
  $\overline{\boldsymbol{c}}^{iter} \leftarrow \sum_{sc} P_{sc} \boldsymbol{c}_{sc}^{iter}$, *Ind*=No. of non-binary elements in $\overline{\boldsymbol{c}}^{iter}$
**Until** $Ind \le \varepsilon$ ($\varepsilon$=0 for TPH, 1 or 2 for MPH)
**If** Apply MPH:
  Locate all the inconsistent UCs $\boldsymbol{c}_{ic}$. Denote other UCs as $\boldsymbol{c}_c$.
  **Parallel For** *case* from 1 to $2^{Ind}$
    Select a distinct combination of $\boldsymbol{c}_{ic}^{case}$.
    $\underset{\boldsymbol{x}_{sc}}{\text{minimize}} \quad OBJ_{case} = \begin{bmatrix} \boldsymbol{\lambda}_{0,c}^T & \boldsymbol{\lambda}_{0,ic}^T \end{bmatrix} \begin{bmatrix} \boldsymbol{c}_c \\ \boldsymbol{c}_{ic}^{case} \end{bmatrix} + \sum_{sc} P^{sc} \boldsymbol{\lambda}_{sc}^T \boldsymbol{x}_{sc}$
    subject to $\left(\boldsymbol{c}_c, \boldsymbol{c}_{ic}^{case}, \boldsymbol{x}_{sc}^{iter}\right) \in \mathcal{Q}_{sc}, \forall sc \in \mathcal{SC}$
  **End Parallel**
  Select the $\boldsymbol{c}_{ic}^{case}$ with minimum $OBJ_{case}$.
**End If**

PH aims to solve each scenario independently, by means of penalizing the differences in each SIV in the optimal solution of each scenario. The common flowchart is outlined in [17] (corresponding to lines 1-15 of **Algorithm 1**), with $\boldsymbol{c}$ as SIVs.

The penalty function $\boldsymbol{\rho}_{sc}^{iter}$ is adjusted iteratively with respect to the current inter-scenario differences, and is added to the objective function until the differences shrink considerably. The penalty stems heuristically from two sources: inter-scenario differences, and a shift in the average UC (respectively corresponding to the terms $\boldsymbol{\rho}_{sc}^{iter-1\,T}\boldsymbol{c}$ and $\frac{\kappa}{2}\left\|\boldsymbol{c}-\overline{\boldsymbol{c}}^{iter-1}\right\|_2^2$ ). The penalty coefficient $\kappa$ is chosen proportional to its objective coefficients as discussed in [17].

The traditional PH algorithm is mostly effective in decreasing the *Ind*, or the number of inconsistent UCs, to a low level. In a medium-scale network where the unit production prices are disparate for different generators, the inconsistency levels easily converge to zero. However, after repeated tests, we discovered that *Ind* may fail to converge to zero when the network is large, or when power generation prices are similar. Instead this indicator remains at 1 or 2 for the next 10+ iterations unless the penalty parameters are carefully chosen for a specific network. Hence, we modified the PH algorithm, by terminating the iteration whenever *Ind* drops below 1 or 2. From then on, all possibilities in inconsistent UCs are enumerated, while the total cost is recomputed with predetermined binaries. The UC set with the lowest objective is selected as the final solution (lines 16-24 in **Algorithm 1**). Compared with TPH, the modified algorithm can effectively shorten the computational time. Only $2^{Ind} \leq 4$ iterations are required to compare the enumerated UCs.

Additional adaptations to the PH algorithm are as follows:

1) In the objective function (1), no coefficients are assigned to the unit commitments $c_{i,t}$ , which poses challenges to determining the penalty function. Regarding this, the objective function is reformulated by expressing the power production $pd_{i,t}^{sc}$ in terms of $c_{i,t}$ , $\underline{P}_i$ and an artificial variable $pd_{i,t}^{sc\prime}$

$$pd_{i,t}^{sc} \rightarrow c_{i,t} \cdot \underline{P}_i + pd_{i,t}^{sc\prime} \tag{22}$$

Each $c_{i,t}$ is thereby related to a unit cost of $C_i^{PD} \cdot \underline{P}_i$ .

2) For continuous variables, PH typically converts the objective function from linear to quadratic as a result of the Euclidean norm $\frac{\kappa}{2}\left\|\boldsymbol{c}-\overline{\boldsymbol{c}}^{iter-1}\right\|_2^2$ . For this model, however, the objective function can retain an MILP format after the relaxation penalty. This is achieved by substituting $c^2$ with $c$, which are equivalent terms for both binary values 0 and 1, within the aforementioned norm. The retained MILP format greatly contributes to a higher calculation efficiency.

### *B. Second Order Cone Relaxation*

To establish a fast and mostly accurate depiction of Constraint (15), an iteration-free second-order cone relaxation is applied. It largely outperforms piecewise linearization in terms of time consumption, due to the exclusion of integer variables. See **Algorithm 2** for the procedures (For simplicity, we denote $CONT = \sqrt{\frac{0.617D_{cd}^5}{L_{cd}FRTZ\rho^2}}$ ).

Similar to previous research on electricity-gas coupled systems [38], the first step is to determine the direction of gas flow *gf* in (15), to remove the absolute sign. The modified equation is then transformed into a convex second-order inequality (line 5 or 9 in **Algorithm 2**) and a concave inequality. While the former inequality is easily solvable, the latter requires advanced techniques to prevent $\left|\pi_{c,t}^{sc\;2} - \pi_{d,t}^{sc\;2}\right|$ from expansion.

**Algorithm 2: Second-Order Cone Relaxation**

**1** Determine the direction of gas flow ($gf > 0$ or $gf \leq 0$?).
**2** **For** each scenario
**3**     **For** each pipeline
**4**         **If** $gf > 0$
**5**             Add $\left\|\begin{matrix} gf_{cd}^{sc} \\ CONT \cdot \pi_{d,t}^{sc} \end{matrix}\right\|_2 \leq CONT \cdot \pi_{c,t}^{sc}$ into the constraint
**6**             Add $gf_{cd}^{sc} \geq CONT \cdot \left(\pi_{c,t}^{sc} - \pi_{d,t}^{sc}\right)$ into the constraint
**7**             Add $\gamma\left(\pi_{c,t}^{sc} - \pi_{d,t}^{sc}\right)$ into the objective function
**8**         **Else** ($gf \leq 0$)
**9**             Add $\left\|\begin{matrix} gf_{cd}^{sc} \\ CONT \cdot \pi_{c,t}^{sc} \end{matrix}\right\|_2 \leq CONT \cdot \pi_{d,t}^{sc}$ into the constraint
**10**             Add $gf_{cd}^{sc} \leq CONT \cdot \left(\pi_{c,t}^{sc} - \pi_{d,t}^{sc}\right)$ into the constraint
**11**             Add $\gamma\left(\pi_{d,t}^{sc} - \pi_{c,t}^{sc}\right)$ into the objective function
**12**         **End If**
**13**     **End For**
**14** **End For**

To boost the computational efficiency, we intentionally discarded iterative constraints to tighten the relaxation. Instead, a penalty function in the form of $\gamma\left(\pi_1 - \pi_2\right)$ is added to the objective (line 7/11 in **Algorithm 2**). In addition, to compensate for over-relaxation from only adopting a rudimentary penalization approach, a set of linear constraints

$$gf \geq (\leq) CONT \cdot \left(\pi_{c,t}^{sc} - \pi_{d,t}^{sc}\right) \tag{23}$$

is imposed (line 6/10). (23) are weaker inequalities of the concave constraint; therefore, the convexity can be restored.

## IV. Case Study

In this section, the IEEE 24-bus power system/Belgium 20-node gas system is used to test the validity of the PH algorithm. We further applied PH to the IEEE 118-bus power system/ Belgium 20-node gas system to study its scalability in IES.

The calculation is performed on a Win10 environment with 8 Intel® Core i7-6700 CPUs, which can simultaneously provide 4 workers for parallel computing. The optimization model is formulated with MATLAB R2016a and Yalmip R20200116 [39], and solved by CPLEX V12.10.0.

### *A. IEEE 24-bus power system/20-node gas system*

In this medium-scale IES, there are 3 gas-fired generators, U1, U4 and U5, connected to the gas system. In addition, there are 5 coal-fired generators (U2, U3, U6, U7 and U8), and 2 wind turbines (U9 and U10). The network parameters are available from [27].

The power and gas loads are considered deterministic. In contrast, the wind power is stochastic. A total of 3000 random windspeed error curves are generated from the ARMA(1,1) series, and reduced to 15 through k-means clustering. After adding error terms to the windspeed baseline, we yield the output curve through power-windspeed relationships.

Two sets of unit production prices are presented to test the performance of the TPH/MPH algorithm:

**Case** 1: Production costs for gas-fired generators are set much lower than those for coal-fired generators. Therefore, aside from renewable energies, gas-fired generators U1, U4 and U5 are prioritized units for power generation.

**Case** 2: Production costs for gas-fired generators are set similar to those for coal-fired generators. The output of each generator is thus more susceptible to the wind and load profiles.

*A1. Numerical Results and Analysis*

Before the stochastic operation of electricity-gas coupled systems is investigated, the optimal result of a separate scenario with higher wind penetration is studied. Fig. 1 depicts the hourly dispatch of three generator types for both cases. In both cases, the gas- and coal-fired outputs fluctuate accordingly with the total demand not met by wind power. However, because of the lower production prices for gas-fired generators in Case 1, an average of 0.128 GW more power is produced from gas. This also leads to gas-fired generators more frequently operating at their full capacity, especially in hours 17 to 23 when the load demand is high.

To illustrate the necessity of considering stochasticity and prove the effectiveness of the proposed stochastic UC model, the UCs determined via the proposed model are compared with those obtained in each scenario separately. Most UCs among different scenarios are the same, with minor exceptions. Table 1 presents two sets of UCs in Case 2 where differences occur.

The first example features the occasional start-up of generator U3 during periods 3-7 and 9-14 in Scenario 1. In the separately optimized model, the high wind power combined with the low power loads forces U3 to be shut down during these periods. However, in hours 3,7,9,10 and 12, the generator needs to start up as a compromise for the other scenarios, where wind power is insufficient. The second set of UCs in Scenario 8 demonstrates the opposite case: U1 needs to be shut down to reduce costs for the other scenarios.

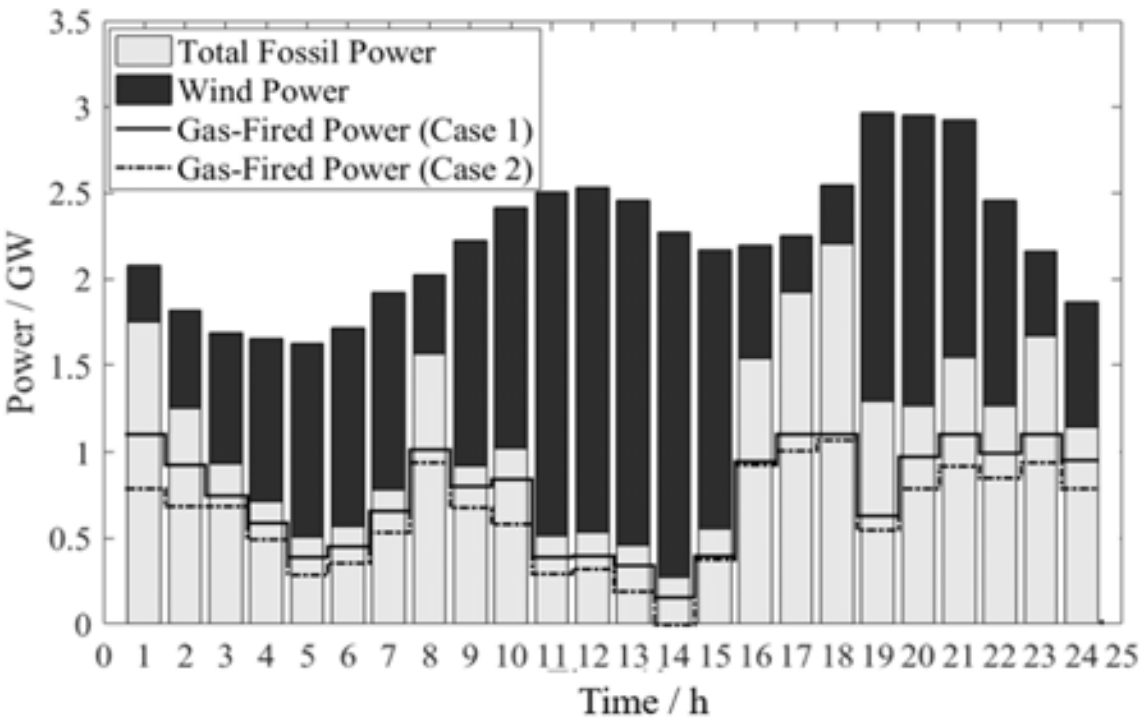

Fig. 1 Hourly dispatch for each generator type in Scenario 1.

Table 1 UC for separate scenarios and the stochastic model (Case 2)

| *Optimization Conditions* | | *Unit Commitment* ● On ○ Off |
|---|---|---|
| Scenario 1, U3 | Separate Scenario | ●●○○○○○●○○○○○○●●●●●●●●●● |
| | MPH-based Stochastic | ●●●○○○●●●●○●○○●●●●●●●●●● |
| Scenario 8, U1 | Separate Scenario | ●●●●●●●●●●●●●●●●●●●●●●●● |
| | MPH-based Stochastic | ●●●●●●●●●●●○○○●●●●●●●●●● |

*A2. Performance of the TPH/MPH algorithm*

The TPH/MPH algorithm unifies the total inconsistent unit commitments with high efficiency. Fig. 5 plots the inconsistency level $\sum_{sc} P_{sc} \left\| \boldsymbol{c}_{sc}^{iter} - \overline{\boldsymbol{c}}^{iter} \right\|_2$ with respect to PH iterations for both cases ($\kappa$ set to 1).

For Case 1, the error converges quickly to zero in 6 iterations. For Case 2, the problem becomes more intricate. Within the first 4 iterations, the inconsistency level converges considerably due to the oriented penalty terms. Iterations 5-9 are a trial-and-error period with a fluctuating inconsistency level. Finally, the error is stabilized (see the zoomed-in figure), where *Ind* remains at 1 for a long period before converging to 0. For TPH, 42 iterations are required before convergence, which is not illustrated in Fig. 2. However, for the proposed MPH, the iterations can be effectively terminated when *Ind* reaches 1 (i.e. iteration 10). The algorithm then searches for the outlier among all UCs, and performs enumeration to determine the best UC for this outlier. The total computational time is equivalent to only 12 iterations in terms of time consumption.

Table 2 Comparison of different cases and methods in the IEEE 24-bus power system/Belgium 20-node gas system.

| *Case* | *Method* | *Computation Time (s)* | *Expected Cost (M$)* | *Coal Production Cost (M$)* | *Gas Production Cost (M$)* | *Non-served Power (GW, Scenario-Averaged)* |
|---|---|---|---|---|---|---|
| ***Case 1*** | Method 1 (benchmark) | 4035 | 14.7299 | 4.7023 | 6.9010 | 0 |
| | Method 2 (Deterministic UC) | 18.74 | 164.90 | 6.7181 | 4.7940 | 0.303 |
| | Method 3 (Benders) | 641.27 | 14.7289 | 4.6604 | 6.9376 | 0 |
| | Method 4 (TPH, 1/2/4 workers) | 65.17/46.24/36.11 | 14.7354 | 4.7192 | 6.8906 | 0 |
| ***Case 2*** | Method 1 (benchmark) | 10933 | 13.9762 | 5.2747 | 5.6602 | 0 |
| | Method 2 (Deterministic UC) | 21.63 | 165.91 | 5.1201 | 5.6977 | 0.307 |
| | Method 3 (Benders) | 407 | 13.9736 | 5.2003 | 5.7358 | 0 |
| | Method 4 (MPH, 1/2/4 workers) | 114.3/70.72/58.03 | 13.9754 | 5.0411 | 5.8752 | 0 |

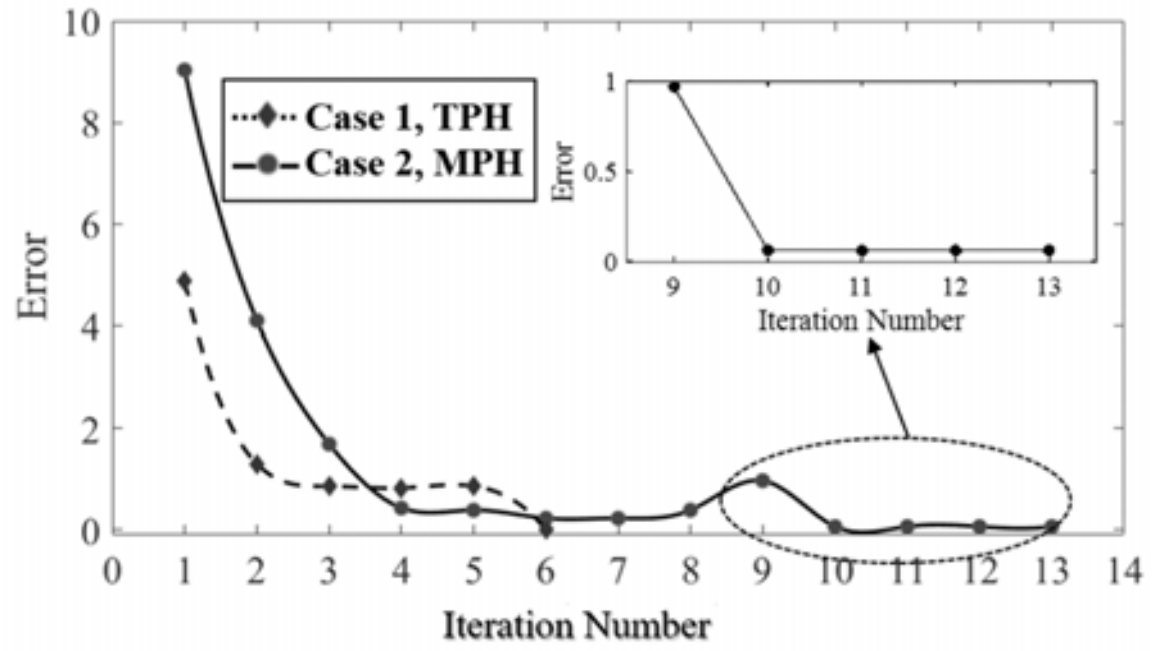

Fig. 2 The convergence of inconsistency levels.

Despite the heuristics applied, Progressive Hedging is still highly competent in obtaining the optimal cost. To illustrate, the following methods are proposed and applied to the model:

**Method 1**: Stochastic UC solved by CPLEX as a benchmark.

**Method 2**: Deterministic UC. A deterministic UC with one scenario is performed, and then the UC result is applied in multi-scenario dispatch to test the effectiveness.

**Method 3**: Stochastic UC solved by Benders Decomposition.

**Method 4**: Stochastic UC solved by TPH/MPH, with parallel computing enabled in MATLAB.

It's noteworthy that we also used piecewise linearization to compare with the proposed iteration-free SOCR for natural gas pipeline modeling. However, it cannot be solved within acceptable time because too many binaries are introduced.

For Case 1, TPH is applied due to the disparate unit pricing, while in Case 2, MPH is applied. The program respectively accesses 1, 2 and 4 workers in the environment, to test the effectiveness of parallel computing in the algorithm.

The time consumption (including the model formulation stage) and expected cost for each method are summarized in Table 2. For the benchmark (CPLEX), difficulties in locating feasible solutions and decreasing the gap led to a total computational time of 4035 s/10933 s. Compared with that of Case 1, the optimization of Case 2 is twice as time-consuming, due to the proximity in unit prices which complicates the dispatch. It can be seen that deterministic UC, despite exhibiting a faster computation speed, is not an ideal alternative to stochastic UC, since a total non-served power of 0.3 GW appears in the final solution.

For the TPH/MPH algorithm, the computational burden is considerably reduced to ~65 s and ~114 s, respectively. The algorithm's support for parallel computing further boosts the computational efficiency. For example, 4-workers in Case 2 shortened the computational time by 49.2% compared to 2-worker servers. The computational speed employing PH even outperforms the Benders Decomposition by at most 17 times.

In Case 1, the UC solutions of TPH and the benchmark method only differ in 1 generation unit. Therefore, the expected cost, as well as the gas-fired and coal-fired power production costs, only differ in 0.01 M$. Interestingly, in Case 2, although the UC solutions differ in 11 generation units, there is little difference in total cost (13.9762 vs 13.9754 M$). The shift in UCs can be explained by similar power production prices, where many solutions are considered optimal with a threshold of $10^{-4}$.

### *B. IEEE 118-bus power system/20-node gas system*

This large-scale IES is utilized to test the scalability of the PH algorithm. The power system is composed of 36 coal-fired generators, 12 gas-fired generators and 6 wind turbines [39]. The gas system is not expanded correspondingly, since the Belgium 20-node gas system is already one of the largest systems in the real world, and investigated most frequently [27][38].

The performance of different methods is presented in Table 3. The PH method can still efficiently obtain the optimal solution. The early termination and enumeration method in MPH are especially useful in such large systems, even when the power production prices are already dissimilar. In this case, MPH significantly reduces the number of iterations needed from 110 to 60. As a result, only 270 s is required to compute the stochastic UC with 4 parallel workers. The total cost only differs from the benchmark by 0.0065%, which again proves the optimality of solution produced by PH.

Benders Decomposition, on the other hand, is inefficient in handling this problem. It is even slower than the benchmark method, which suggests that its efficiency might be heavily subject to the scale and parameters of the network.

Table 3 Comparison of different cases and methods in the IEEE 118-bus power system/Belgium 20-node gas system.

| *Method* | Computational Time (s) | Expected Cost (M$) |
|---|---|---|
| benchmark | 1227.0 | 1.5425 |
| Benders | 1962.5 | 1.5429 |
| TPH<br>1/2/4 workers | 802.8/544.7/514.8 | 1.5426 |
| MPH<br>1/2/4 workers | 406.3/285.5/270.0 | 1.5426 |

## V. Conclusion

In this article, we proposed a stochastic UC model for electric-gas coupled IESs. Moreover, we applied and modified the PH algorithm to accelerate the optimization of the stochastic UC. It was proven by case studies that, for a medium-scale electric-gas coupled IES, TPH easily converges when generators have disparate unit production prices, while for similar unit pricing, MPH is required for convergence via early termination and UC enumeration. For large-scale IES, MPH is inherently useful for saving computational time, regardless of generation prices. Combined with an iteration-free SOCR method to restore the convexity of the IES, the inconsistency level among scenarios converges quickly, and the computational speed of the proposed algorithm is at most 188 times faster than that of commercial software. The acceleration is attributable to the uniform Decomposition of the stochastic model, and the algorithm's support for parallel computing. Meanwhile, the gap between PH algorithm and the benchmark is less than 0.01% in both IES systems, which proves that the solution produced by PH reaches acceptable optimality in this stochastic UC problem.

The TPH/MPH algorithm might also help solve stochastic UC problems in other IESs, e.g. electricity-heat coupled system. We will further explore this topic in the future.

**References**

[1] EIA, Annual Energy Review, 2011 [Online]. Available: http://www.eia.doe.gov/emeu/aer.

[2] C. M. Correa-Posada and P. SaNchez-Martin, and I. N. Sneddon, "Security-constrained optimal power and natural-gas flow," in IEEE Trans. Power Syst., vol. 29, pp. 1780-1787, 2014.

[3] R. E. Robert, M. P. Croissant and J R. Masih, "International energy outlook: U. S. department of energy," in Washington Quarterly, vol. 19, pp. 70-99, 1996.

[4] IGU. IGU World LNG Report – 2017 Edition [Online]. Available: http://www.igu.org/sites/default/files/node-document-field_file/103419-World_IGU_Report_no%20crops.pdf.

[5] M. Gotz, J. Lefebvre, F. Mors, et al., "Renewable Power-to-Gas: A technological and economic review," in Renewable Energy, vol. 85, pp. 1371-1390, 2016.

[6] A. Alabdulwahab, A. Abusorrah, X. Zhang and M. Shahidehpour, "Coordination of Interdependent Natural Gas and Electricity Infrastructures for Firming the Variability of Wind Energy in Stochastic Day-Ahead Scheduling," in IEEE Transactions on Sustainable Energy, vol. 6, no. 2, pp. 606-615, April 2015.

[7] X. Zhang, M. Shahidehpour, A. Alabdulwahab and A. Abusorrah, "Hourly Electricity Demand Response in the Stochastic Day-Ahead Scheduling of Coordinated Electricity and Natural Gas Networks," in IEEE Transactions on Power Systems, vol. 31, no. 1, pp. 592-601, 2016.

[8] C. He, L. Wu, T. Liu and Z. Bie, "Robust Co-Optimization Planning of Interdependent Electricity and Natural Gas Systems With a Joint N-1 and Probabilistic Reliability Criterion," in IEEE Transactions on Power Systems, vol. 33, no. 2, pp. 2140-2154, March 2018.

[9] C. Wang, W. Wei, J. Wang et al., “Robust Defense Strategy for Gas-Electric Systems Against Malicious Attacks” IEEE Trans. Power Syst., vol. 32, pp. 2953-2965, 2016.

[10] Z. Qiao, Q. Guo, H. Sun, Z. Pan, Y. Liu and W. Xiong, “An interval gas flow analysis in natural gas and electricity coupled networks considering the uncertainty of wind power” Applied Energy 201, pp. 343-353, 2017.

[11] A. Alabdulwahab, A. Abusorrah, X. Zhang and M. Shahidehpour, "Stochastic Security-Constrained Scheduling of Coordinated Electricity and Natural Gas Infrastructures," in IEEE Systems Journal, vol. 11, no. 3, pp. 1674-1683, Sept. 2017.

[12] A. Ben-Tal and A. Nemirovski, "Robust optimization – methodology and applications," in Mathematical Programming, vol. 92, no. 3, pp. 453-480, May 2002.

[13] T. Li, M. Shahidehpour and Z. Li, "Risk-Constrained Bidding Strategy With Stochastic Unit Commitment," in IEEE Transactions on Power Systems, vol. 22, no. 1, pp. 449-458, Feb. 2007.

[14] L. Wu, M. Shahidehpour and T. Li, "Cost of Reliability Analysis Based on Stochastic Unit Commitment," in IEEE Transactions on Power Systems, vol. 23, no. 3, pp. 1364-1374, Aug. 2008.

[15] R. Egging, "Benders Decomposition for multi-stage stochastic mixed complementarity problems – Applied to a global natural gas market model, " in European Journal of Operational Research, vol. 226, no. 2, pp. 341-353, Apr 2013.

[16] R. T. Rockafellar and RJ.-B. Wets, "Scenarios and policy aggregation in optimization under uncertainty," in Math Oper Res, vol. 16, pp. 119–147, 1991.

[17] J.-P. Watson and D. L. Woordruff, "Progressive hedging innovations for a class of stochastic mixed-integer resource allocation problems," in Comput Manag Sci, vol. 8, pp. 355-370, 2011.

[18] S. Arpón, T. Homem-de-Mello and B. K. Pagnoncelli, "An ADMM algorithm for two-stage stochastic programming problems," in Ann Oper Res, vol. 286, pp. 559–582, 2020.

[19] J. Jian, C. Zhang, L. Yang and K. Meng, "A hierarchical alternating direction method of multipliers for fully distributed unit commitment," in International Journal of Electrical Power & Energy Systems, vol. 108, pp. 204-217, 2019.

[20] S. M. Ryan, R. J. -. Wets, D. L. Woodruff, C. Silva-Monroy and J. Watson, "Toward scalable, parallel progressive hedging for stochastic unit commitment," 2013 IEEE Power & Energy Society General Meeting, Vancouver, BC, 2013, pp. 1-5.

[21] H. Wu et al., "Stochastic Multi-Timescale Power System Operations With Variable Wind Generation," in IEEE Transactions on Power Systems, vol. 32, no. 5, pp. 3325-3337, Sept. 2017.

[22] D. Gade, G. Hackebeil, S. M. Ryan et al., “Obtaining lower bounds from the progressive hedging algorithm for stochastic mixed-integer programs,” in Math. Program., vol. 157, pp. 47–67, 2016.

[23] A. M. Palani, H. Wu and M. M. Morcos, "A Frank–Wolfe Progressive Hedging Algorithm for Improved Lower Bounds in Stochastic SCUC," in IEEE Access, vol. 7, pp. 99398-99406, 2019.

[24] J. Qiu, H. Yang, Z. Y. Dong et al., “A Linear Programming Approach to Expansion Co-planning in Gas and Electricity Markets,” in IEEE Trans. Power Syst., vol. 31, pp. 3594-3606, 2015.

[25] Seungwon An, Qing Li and T. W. Gedra, "Natural gas and electricity optimal power flow," 2003 IEEE PES Transmission and Distribution Conference and Exposition (IEEE Cat. No.03CH37495), Dallas, TX, USA, 2003, pp. 138-143 Vol.1.

[26] C. Liu, M. Shahidehpour and J. Wang, "Coordinated scheduling of electricity and natural gas infrastructures with a transient model for natural gas flow, " in CHAOS, vol. 21, pp. 025102, 2011.

[27] C. M. Correa-Posada and P. Sanchez-Martin, "Integrated Power and Natural Gas Model for Energy Adequacy in Short-Term Operation," in IEEE Trans. Power Syst., vol. 30, pp. 3347-3355, 2015.

[28] C. M. Correa-Posada and P. Sánchez-Martín, "Gas network optimization: A comparison of piecewise linear models." in Optimization Online, 2014.

[29] L. Yang, Y. Xu, H. Sun and X. Zhao, "Two-stage Convexification Based Optimal Electricity-Gas Flow," in IEEE T SMART GRID.

[30] Y. He et al., "Decentralized Optimization of Multi-Area Electricity-Natural Gas Flows Based on Cone Reformulation," in IEEE Transactions on Power Systems, vol. 33, no. 4, pp. 4531-4542, 2018.

[31] S. Chen, A. J. Conejo, R. Sioshansi and Z. Wei, "Unit Commitment With an Enhanced Natural Gas-Flow Model," in IEEE Transactions on Power Systems, vol. 34, no. 5, pp. 3729-3738, Sept. 2019.

[32] A. Schwele, C. Ordoudis, J. Kazempour, P. Pinson, “Coordination of power and natural gas systems: Convexification approaches for linepack modeling,” in Proceedings of IEEE PES PowerTech 2019. Milan, Italy.

[33] A. Belderbos, K. Bruninx, E. Delarue and W. D'haeseleer,"Facilitating Renewables and Power-to-gas via Integrated Electrical Power-gas System Scheduling." Applied Energy, 2020.

[34] F. Liu, Z. Bie, X. Wang, “Day-ahead dispatch of integrated electricity and natural gas system considering reserve scheduling and renewable uncertainties,” in IEEE Transactions on Sustainable Energy vol. 10, no. 2, pp. 646–658.

[35] NDBC Standard Meteorological Buoy Data, 1970-present. URL: coastwatch.pfeg.noaa.gov/erddap/tabledap/cwwcNDBCMet.html.

[36] L. Soder, "Simulation of wind speed forecast errors for operation planning of multiarea power systems," 2004 International Conference on Probabilistic Methods Applied to Power Systems, Ames, IA, pp. 723-728.

[37] J. A. Carta, "Wind Power Integration," in Comprehensive Renewable Energy, vol. 2, 2012, pp. 569-655.

[38] C. Wang, W. Wei, J. Wang, L. Bai, Y. Liang and T. Bi, "Convex Optimization Based Distributed Optimal Gas-Power Flow Calculation," in IEEE Transactions on Sustainable Energy, vol. 9, no. 3, pp. 1145-1156, July 2018.

[39] C. M. Correa-Posada and P. Sánchez-Martín, "Security-constrained unit commitment with dynamic gas constraints," 2015 IEEE Power & Energy Society General Meeting, Denver, CO, 2015, pp. 1-5.

[40] J. Lofberg, "Yalmip: A toolbox for modelling and optimziation in MATLAB," in Proc. CASCD, Taipei, 2004, pp. 284-289.